\documentstyle[aps,prl,epsf]{revtex} 

\begin{document} 
\draft 
\twocolumn[\hsize\textwidth\columnwidth\hsize\csname 
@twocolumnfalse\endcsname 
\title{Measuring Condensate Fraction in Superconductors} 
\author{Sudip Chakravarty and Hae-Young Kee} 
\address{Department of Physics and Astronomy, University of 
California 
Los Angeles\\ 
Los Angeles, CA 90095-1547} 
\date{\today} 
\maketitle 
\begin{abstract} 
An analysis of  off-diagonal long-range order in superconductors shows that the
spin-spin correlation function is significantly influenced by the order if the
order parameter is anisotropic on a microscopic scale. Thus, magnetic  neutron
scattering  can provide a direct measurement of the condensate fraction of a
superconductor. It is also argued that recent measurements in high temperature
superconductors come very close to achieving this goal.
\end{abstract} 
] 

Off-diagonal long-range order (ODLRO)\cite{Penrose,Yang} provides an intriguing   
characterization of superconductors and superfluids for which the order parameter
signifies  a unique  coherence property of a macroscopic quantum system
corresponding to  the spontaneous breakdown of global gauge symmetry. It is
therefore unfortunate  that direct measurements of this order parameter are
few and far between\cite{exptl}.  

For BCS superconductors, all definitions of superconductivity are directly linked
to the  energy gap in the single particle excitation spectrum. Therefore, the
observation of the gap, which is abundant,   is virtually the same as the
observation of ODLRO. In contrast,  the high temperature superconductors
sometimes exhibit an apparent gap at
temperatures well above the superconducting transition temperature, $T_c$, similar
to the gap seen below
$T_c$\cite{Randeria}. In those instances, the presence of a  gap is  not the
same as the presence of ODLRO, and it  becomes  necessary to explore ODLRO in
more general terms, not  restricted by  the BCS theory.

The purpose of the present Letter is to offer some insight into the question
of ODLRO in high temperature superconductors and to indicate experiments that may 
shed light on this topic. In fact, we argue that  recent magnetic neutron
scattering measurements come tantalizingly close to accomplishing
this goal\cite{Neutron}.  We show that the instantaneous spin-spin
correlation function is   influenced by the existence of ODLRO. This is partially because the
spin operators are composite Fermion operators, and the correlation function can
acquire anomalous expectation value, but also because the anisotropic order parameter strongly enhances
the effect. We also deduce that strong intercation effects are necessary for the magnitude of the
effect to be observable. 

\paragraph{Summary of experiments:}We begin with a brief summary of experiments.
The magnetic neutron scattering cross section is proportional to the
equilibrium dynamic structure factor, ${\cal S}({\bf k},\omega)$.
The corresponding  instantaneous structure factor,
${\cal S}({\bf k})=\int_{-\infty}^{\infty}d\omega {\mathcal S}({\bf
k},\omega)$, is
\begin{equation}
{\mathcal S}({\bf k})=v_0\sum_{\bf l}e^{i\bf{k\cdot
R_l}}\langle {\bf S}(0)\cdot{\bf S}(\bf{l})\rangle,\label{sk_def}
\end{equation}
where $v_0$ is the volume of the unit cell, and $\bf S$ is the spin operator.

For a class of bilayer cuprate superconductors,
YBa$_2$Cu$_3$O$_{6+\delta}$ and Bi$_2$Sr$_2$CaCu$_2$O$_{8+\delta}$,  corresponding to 
optimal doping, the
${\mathcal S}({\bf k},\omega)$ has a resolution limited resonance peak,
at an energy $\approx 40$ meV and the special wavevector ${\bf k=q}=(\pi/a,\pi/a,\pi/c)$,
for
$T<T_c$; $a$ and $c$ are the lattice constants. More precisely, the ${\mathcal S}({\bf
k},\omega)$ can be classified into parts that are even and odd with respect to the individual
layers  of a bilayer. The resonance appears in the odd channel ($ q_z=\pi/c$), and its
energy is almost independent of temperature, but its intensity vanishes for
$T>T_c$. For underdoped superconductors, there is also a peak for $T>T_c$, but its character is
quite different, as we comment below.

The existence of the resonance peak below $T_c$ allows us to isolate a
distinct {\em part} of the instantaneous structure factor by integrating with
respect to frequency across the resonance, which has a simple theoretical
interpretation in terms of the condensate fraction, as we shall see. In this sense,
the resonance peak plays an essential role; in its absence, it would be difficult
to obtain the same information, as in  La$_{2-\delta}$Sr$_\delta$CuO$_4$.

\paragraph{The Off-diagonal long-range order:} The ODLRO for a singlet
superconductor is defined as a property of the two particle density matrix, which
is
\begin{equation}
\rho^{(2)}( 1\uparrow,
2\downarrow;3\uparrow,4\downarrow)\equiv\langle\psi_{\uparrow}^{\dagger}({\bf
x}_1)\psi_{\downarrow}^{\dagger}({\bf x}_2)\psi_{\downarrow}({\bf
x}_4)\psi_{\uparrow}({\bf x}_3)\rangle
\end{equation}
where $ \psi_{\uparrow}^{\dagger}(\bf x)$ is the  creation
operator of the Fermion field, of spin up and location $\bf x$. Its existence implies that
\begin{eqnarray}
|\rho^{(2)}( 1\uparrow,2\downarrow ;3\uparrow,4\downarrow) - \Phi({\bf
x}_1\uparrow, {\bf x}_2\downarrow)^*\Phi({\bf x}_3\uparrow, {\bf
x}_4\downarrow)|\nonumber\\
\le  (\frac{N}{V})\gamma(|{\bf x}_1-{\bf x}_3|,|{\bf
x}_2-{\bf x}_4| ),
\end{eqnarray}
where  the the non-negative function
$\gamma (r_1,r_2) $, which is independent of
$N$, tends to zero if either $r_1$ or $r_2$ tends to $\infty$. 

Following Ref.~\cite{Penrose}, one can  show that in the thermodynamic limit, the number of
particles  $N\to \infty$ and the volume $V\to \infty$ such that $(N/V)\to n$, the function 
$\Phi$ is the  eigenfunction of the two particle density matrix
corresponding to the largest
eigenvalue $\lambda_M$\cite{eigenvalue}, given by
\begin{equation}
\lambda_M=\int d{\bf x}_1d{\bf x}_2|\Phi({\bf
x}_1\uparrow,{\bf x}_2\downarrow)|^2,
\end{equation}
where $\lambda_M/N$ is the condensate fraction defined to be  the fraction of electrons
participating in pairing, 
not to be mistaken to be the superfluid density, $n_s$, which is  the  stiffness with
respect to an imposed twist of the order parameter.
Hence, in the thermodynamic limit, the following spectral
decomposition of the density matrix holds for {\em all} separations:
\begin{eqnarray}
\rho^{(2)}( 1\uparrow,2\downarrow;3\uparrow,4\downarrow) &=&
\Phi({\bf x}_1\uparrow,
{\bf x}_2\downarrow)^*\Phi({\bf x}_3\uparrow, {\bf
x}_4\downarrow)\nonumber \\ &+&
G({\bf x}_1\uparrow,{\bf x}_2\downarrow;
{\bf x}_3\uparrow,{\bf x}_4\downarrow).\label{spectral_decomposition}
\end{eqnarray}
The function
$G$  vanishes when the
separation between the  groups
$({\bf x}_1,{\bf x}_2)$ and $({\bf x}_3,{\bf x}_4)$
tends to $\infty$. While the first term of Eq.~(\ref{spectral_decomposition})
vanishes above $T_c$, the second term continues smoothly to the normal state
above $T_c$. 

For a translationally invariant system corresponding to
a condensate of zero total momentum, we can write
$
\Phi({\bf x}_1\uparrow,{\bf x}_2\downarrow)\equiv \Phi({\bf x}_1-{\bf
x}_2)\left(|\uparrow\downarrow\rangle-
|\downarrow\uparrow\rangle\right)/\sqrt{2}
$,
where $\Phi({\bf x})=\Phi(-{\bf x})$. 

For a BCS superconductor (in an abbreviated notation),
\begin{equation}
\rho^{(2)}=\Phi_0({\bf x}_{12})^*\Phi_0({\bf x}_{34})+g_0({\bf
x}_{13})^*g_0({\bf x}_{24}),
\end{equation}
where the BCS function $\Phi_0({\bf x}_{12})\equiv \Phi_0({\bf x}_1-{\bf x}_2)$ is a
function of the separation between the mates of a Cooper pair. Its integral $(1/N)\int d{\bf x}
|\Phi_0({\bf x})|^2$ is the condensate fraction. In BCS theory, and in the absence of
disorder, $n_s=n$, while the condensate fraction is
$\approx N(0)\Delta$, where $\Delta$ is the gap, and $N(0)$ is the density of states at
the Fermi energy. As $T\to T_c$, the condensate fraction tends to zero as
$\Delta^2N(0)/T_c$, similar to $n_s$. In 
general,  these are distinct concepts, however.

The BCS function $g_0(\bf x)$ is the parallel spin correlation function modified by
the presence of  a gap, which smoothly continues to the normal state. In contrast,
the function $\Phi_0(\bf x)$ vanishes in the normal state.

The discussion above can be summarized as follows: (1) the spectral decomposition
in Eq. (\ref{spectral_decomposition}) is general and does not depend on the BCS
theory; (2) the order parameter function $\Phi$ includes, in principle, all
effects of electron-electron interaction; (3) in common with BCS theory, 
$\lambda_M$ vanishes above $T_c$, while the contribution due to $G$ continues
smoothly above
$T_c$; (4) as in BCS theory, the function $G$ must be a global gauge singlet.

\paragraph{The spin-spin correlation function:} From here on we shall abandon the
continuum notation and adopt the more  natural
lattice notation. Assuming that the spin rotational symmetry is unbroken, we note
that 
\begin{eqnarray}
\langle {\bf S}({\bf l}_1)\cdot{\bf S}({\bf l}_2)\rangle
=\frac{3}{4}\delta_{1,2}n_o-\frac{3}{2}\rho^{(2)}(1\uparrow,
2\downarrow;2\uparrow,1\downarrow),
\end{eqnarray}
where $n_o$ is the average occupation of a site. 
Making use of the spectral decomposition of the two-particle density matrix, we
can write
\begin{eqnarray}
\langle {\bf S}({\bf l}_1)\cdot{\bf S}({\bf l}_2)\rangle&=&\frac{3}{4}
\delta_{{\bf l}_1,{\bf l}_2} n_o 
-\frac{3}{2}[
\Phi({\bf l}_1\uparrow,
{\bf l}_2\downarrow)^*\Phi( {\bf l}_2\uparrow, {\bf l}_1\downarrow)\nonumber \\
&+&G({\bf l}_1\uparrow,{\bf l}_2\downarrow;{\bf l}_2\uparrow,{\bf
l}_1\downarrow)].\label{spin_spin}
\end{eqnarray}

Then remembering
spin rotational symmetry and using Eqs.~\ref{spin_spin} and \ref{sk_def}, we get
\begin{equation}
{\mathcal S}({\bf k})
=\frac{3}{4}v_0n_0-\frac{3}{2}v_0\sum_{{\bf l}}e^{i\bf{k\cdot
R_l}}[\lambda_M |f({\bf l})|^2+G({\bf l})],\label{sk}
\end{equation}
where $f(\bf l)$ is now assumed to be normalized ($v_0\sum_{\bf l}|f({\bf l})|^2=1$). From the
perspective of ODLRO, the summation variable
$\bf l$ in Eq.~(\ref{sk}) is a degree of freedom internal to the Cooper pair, denoting the
separation between its mates. In the absence of simultaneous diagonal
long range order, $G(\bf l)$ is a short ranged quantity, which is 
qualitatively unaffected by  superconductivity, essentially because the short-distance 
electron-electron interaction  is similar in both the normal and the
superconducting phases. The underdoped cuprates,
discussed below, are more complex.

We first consider a single CuO-plane,
ignoring the fluctuations that  destroy ODLRO at any finite temperature,
$T$;  the coupling between the planes is  discussed below in section $(f)$. The
wavevectors $\bf k$ and $\bf q$ are to be interpreted as two-dimensional ($2D$) vectors, strictly
${\bf k}_{\parallel}$ and ${\bf q}_{\parallel}$, until the section $(f)$; similarly, ${\bf
R}_{\bf l}$ is to be interpreted as a $2D$ vector. We hope that this is not a cause
for confusion.

\paragraph{Optimally doped superconductors:} Consider first the case of optimally
doped superconductors for which 
$\mathcal S (\bf k)$ is peaked at
$\bf k = q$ for $T<T_c$, but not so for $T>T_c$. Here, $\bf q$ is the
vector $(\pi/a,\pi/a)$. In Eq.~(\ref{sk}), the exponential factor changes sign
rapidly on the scale of a lattice spacing when
$\bf k = q$ . For any  functions   $|f({\bf
l})|^2$ and
$G({\bf l})$ that are smooth on the scale of a lattice spacing, the sum will be
negligible. 

An order parameter function $f({\bf l})$ such that $|f({\bf
l})|^2$  vanishes when
the vector $\bf R_l$ connects sites  belonging to the same
sublattice of a bipartite square lattice\cite{Henley} can lead to a peak in
$\mathcal S(\bf k)$, because
\begin{equation}
\sum_{\bf l} e^{i\bf q\cdot\bf R_l}|f({\bf l})|^2=-\sum_{{\bf l}\in A}|f(\bf
l)|^2,
\end{equation}
where the sum on the right hand side is over the sublattice $A$. Clearly, a
$d$-wave order parameter satisfies this criterion, but so does
an anisotrpic $s$-wave order. The
distinction between them can only be made on energetic grounds. The
range of $f(\bf l)$ is the ``size of the Cooper pair".

From the argument given above, we expect $G(\bf l)$ to be largely unaffected by
superconductivity. We can then  focus on the quantity
\begin{equation}
\Delta {\mathcal S}({\bf k+q})=\frac{3}{2}v_0\lambda_M \sum_{{\bf
l}\in A}e^{i\bf{k\cdot R_l}}|f({\bf l})|^2.
\end{equation}
Then, noting that $f(\bf l)$ is normalized, we get
\begin{equation}
\frac{\Delta {\mathcal S}({\bf q})}{N}=\frac{3}{2}\frac{\lambda_M}{N}. 
\end{equation}
The inverse $\bf k$-width of  $\Delta\mathcal S(\bf k+
q)$ is the size of the Cooper pair. This result is consistent with a general inequality  derived
in Ref.~\cite{Demler}. The authors of Ref.~\cite{Demler} have
further approximated the inequality to obtain  $\Delta{\cal S}({\bf q})\ge (3/2) |\Delta|^2/x$,
where
$\Delta$ is the  nearest-neighbor $d$-wave order parameter, and $x$ is the concentration of holes. 

If $\lambda_M$ were of the magnitude
predicted for a BCS superconductor, that is, $N(0)\Delta$, the intensity would be
below the sensitivity limit of current neutron scattering measurements. That
the intensity is readily detectable means that the condensate fraction in the
high temperature superconductors is more than an order of magnitude
greater than that predicted in BCS theory. For tightly bound noninteracting Cooper pairs,
acting as  molecules, the condensate density would be the full density $n$.

\paragraph{Underdoped superconductors:} For underdoped superconductors\cite{Fong}, $\cal
S(\bf k)$ is peaked above $T_c$ at the wavevector $\bf q$, but  the growth of the
intensity is  gradual,  increasing by about 20\%
from 300 K to $T_c \sim 50$ K, seen in  both  even and odd channels.  In contrast, the
intensity in the odd channel rises by almost a factor of 2 below
$T_c$. It is very natural therefore to conclude that $G(\bf
l)$ is of very different character than $f(\bf l)$.
Thus, even in the underdoped case, it is possible to separate out the effect of
$G(\bf l)$ and to focus on the contribution solely due to $f(\bf l)$. Before interpreting the data in
terms of  condensate fraction, we remark briefly on the peaked nature of ${\cal S}(\bf k)$ above
$T_c$; there are a number of distinct  possibilities. 

The first possibility is that the spin correlations are described by a
quantum disordered state for which the structure factor is peaked at the
commensurate wavevector $\bf q$. For such a quantum disordered state, it is
difficult, however, to construct a
$d$-wave modulation of the  gap in the single particle excitation spectrum as seen in photoemission
experiments\cite{Hanke}.

The second possibility is  superconducting fluctuations above $T_c$. While such
fluctuations must exist above $T_c$, especially in the underdoped regime, it is unlikely that
they produce a peaked  $\cal S(\bf k)$ at temperatures as high as 300 K. 

The third possibility is a flux state for which the saddle point result\cite{Marston} for the
insulator  is remarkably isomorphic to the BCS result for a superconductor, and  ${\cal S}(\bf k)$ is
peaked at the wavevector $\bf q$, for essentially the same reason. Unfortunately,
it is difficult to see why this should hold as the system is doped with holes,
although it has been argued\cite{Laughlin} that the
notions of the flux state should  continue to hold for the doped case. 

\begin{figure}[htb]
\centering
\epsfxsize=\linewidth
\epsffile{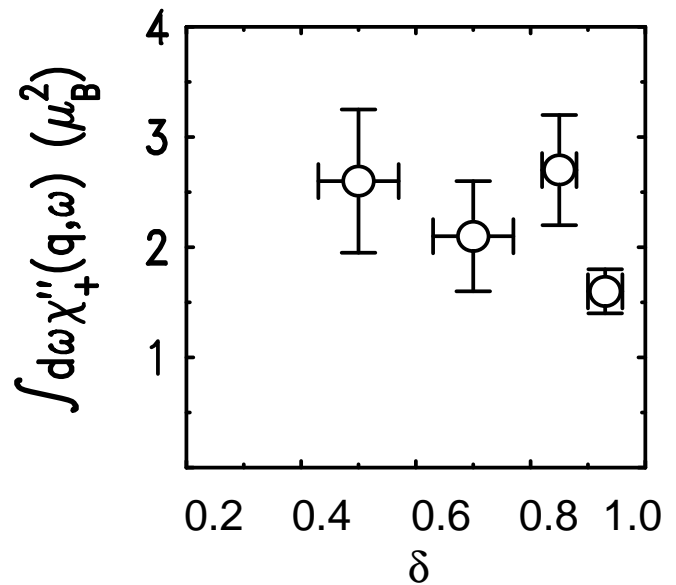}
\caption{The  imaginary part of the dynamical
susceptibility, $\chi_+''(q,\omega)$ of YBa$_2$Cu$_3$O$_{6+\delta}$ at 10 K [9]. This is  proportional 
to
$\Delta {\cal S}(\bf q)$, because, at such low temperatures, the difference
between the two is negligible. The
subscript
$+$ refers to the enhancement over the normal state close to $T_c$; the data is for the odd channel,
as defined in the text. The parameter $\delta\approx 0.42$ corresponds to the crtical
hole concentration $x_c$ at  which the superconductivity disappears at $T=0$.}
\label{inten}
\end{figure}

The interpretation of $\Delta {\cal S}(\bf q)$, in terms of the condensate
fraction has interesting consequences for the quantum phase transition as a function of the hole
concentration,
$x$, hence for the global phase diagram. As 
$x\to x_c^+$, typically $\sim$5\%, the superfluid
density is known to vanish\cite{Batlogg}, but the condensate fraction  increases as shown in
Fig.~\ref{inten}. This is unusual and calls for an explanation.  

A possible resolution is that the experimental system cannot be considered pure, at least for $x$ close
to
$x_c$, and  that this transition is radically altered by disorder, allowing the condensate fraction
to remain finite at the transition. The discontinuous drop
in the condensate fraction at $x_c$ may suggest a first order transition, but it need not be. There
is at least one example in statistical mechanics where it is known that the order parameter vanishes
discontinuously, yet the correlation length diverges at the transition\cite{Ising}. The role of
disorder at this quantum phase transition may then be of crucial importance, and any discussion of a
quantum critical point\cite{Laughlin2} must be reconsidered in this light.

There may be another possible resolution\cite{Chakravarty}. The pseudogap phase may be  a discrete
broken symmetry phase in disguise, known variously as ``flux'' or ``dimer'' ordered state, which has
a peaked ${\cal S}(\bf k)$.  Superconductivity  then develops on top of this. 
Such a broken symmetry state  is easily corrupted by disorder, notoriously plentiful in this
class of materials, which is why it has been identified experimentally
only as a ``crossover'' in some materials and not at all in others.
This crossover will sharpen to an actual phase transition
as the material qualities improve. The neutron intensity within the resonance peak is then a
superposition of two effects, one due to the condensate fraction and the other the due to the discrete
broken symmetry.

We cannot, of course, rule out the possibility that future measurements may reveal that the
condensate fraction turns around and vanishes continuously at
$x_c$, especially because it is so difficult to obtain the absolute magnitude of the intensity in the
neutron measurements. If this is the case, the interpretation of the quantum critical point at
$x_c$ will be straightforward. There is some evidence that the intensity may  vanish at $x_c$. The
data shown in Fig.~\ref{inten} was obtained by integrating over only the {\em positive} part of
the imaginary part of the  susceptibility enhanced over the normal state.  Thus,  a more complete
integration over the frequency will be   necessary to obtain the true theoretical value of the
condensate fraction.

\paragraph{Three-dimensional (3D) and bilayer couplings}
In bilayer  superconductors, the adjacent layers are
grouped into CuO-bilayers, which are then very weakly coupled. The  small coupling
between the bilayers should have a small effect on the $2D$ spin
fluctuations seen above $T_c$, and indeed this is consistent with experiments.
Likewise, such small couplings should have small effects also on the ordered
state at
$T=0$. Close to $T_c$, small $3D$ couplings should  affect the
critical behavior, however, and, below $T_c$, these  couplings are necessary to establish an
order parameter. As far as these weak $3D$ couplings of groups of bilayers are concerned,
qualitatively, little needs to be added to the previous discussion. These comments would
be hardly modified if we were considering superconductors in which single layers (instead of
groups of bilayers) were coupled by weak $3D$ coupling.

In contrast, there is  evidence that the individual layers within a bilayer are
reasonably strongly coupled, thereby splitting the bilayer spin responses that are odd and
even with respect to the individual layers. Although the dynamics is beyond the scope of the
present paper,  we can make a few brief qualitative remarks. For
magnetic neutron measurements, the relevant coupling must be   an antiferromagnetic
superexchange, which must gap the even response function at low frequencies, as in experiments. 
This can be seen by noting that the spin correlations are approximately described by a bilayer
$\sigma$-model\cite{Yin} and expanding the  Euclidean Lagrangian to quadratic
order\cite{footnote}. The magnitude of the bilayer gap is a function of the
$\sigma$-model parameters and must depend on doping, as in experiments\cite{Fong}. As a  result,
from the
$3D$ dynamic structure factor, $S^{ 3\rm D}({ \bf k}, \omega)= \sin^2({ k_z
c \over 2}) S_{ \rm a}^{ 2\rm D}({\bf k}_{\parallel},\omega)+
\cos^2({ k_z c \over 2}) S_{ \rm s}^{ 2\rm D}({\bf k}_{\parallel}, \omega)$, the
symmetric (even) part can be dropped at low frequencies. The integration across the
resonance, which appears sharply in the odd (antisymmetric) channel, then gives the
required structure factor.

In conclusion, we have investigated the static
structure factor for which we could make precise statements, although the
experiments on this are far from complete. We hope that future
experiments will shed further light on the  questions regarding the
condensate fraction  raised here. It remains to be seen if similar precise statements could be
made for the dynamical structure factor as well.  

We thank E. Abrahams, R. J. Birgeneau, P. Bourges, R. B. Laughlin, B. Keimer, C.
Nayak, and S. -C. Zhang for comments. S. C. acknowledges the
grant NSF-DMR-9971138. The work of H. -Y. K  was conducted under the auspices of the Department
of Energy, supported (in part) by funds provided by the University of California for the conduct
of discretionary research by Los Alamos National Laboratory. A part of this work was carried out
at the Aspen Center for Physics.


\begin{references} 
\bibitem{Penrose}O. Penrose and L. Onsagar, Phys. Rev. {\bf 104}, 576 (1956).
\bibitem{Yang}C. N. Yang, Rev. Mod. Phys. {\bf 34}, 694 (1962).
\bibitem{exptl}For a
recent review on condensate fraction in $^4$He, see P. Sokol in {\em Bose-Einstein Condensation},
edited by A. Griffin, D. W. Snoke, and S. Stringari (Cambridge University Press, Cambridge,
1995).
\bibitem{Randeria}For a brief review, see, for example, M.
Randeria, Varenna Lectures, 1997 (cond-mat/9711232).
\bibitem{Neutron}H. F. Fong {\em et al.},
Phys. Rev. Lett. {\bf 75}, 316 (1995); P. Bourges {\em et al.}, Phys. Rev. B {\bf 53},
876 (1996); H. F. Fong {\em et al.}, Phys. Rev. B {\bf 54}, 6708 (1996); P. Dai {\em et
al.}, Phys. Rev. Lett. {\bf 77}, 5425 (1996); H. F. Fong {\em et al.}, Phys. Rev. Lett.
{\bf 78}, 713 (1997); P. Bourges {\em et al.}, Europhys. Lett. {\bf 38}, 313 (1997); H.
A. Mook {\em et al.}, Nature {\bf 395}, 580 (1998); H. F. Fong {\em et al.}, Nature {\bf
398}, 588 (1999).
\bibitem{eigenvalue}The  eigenvalue $\lambda_M$ is  assumed to be nondegenerate because  these
superconductors  are not in a superposition of macroscopically distinct  ODLRO states.
\bibitem{Henley}C. L. Henley, Phys. Rev. Lett. {\bf 80}, 3590 (1998).
\bibitem{Demler} E. Demler, H. Kohno, and S. -C. Zhang, Phys. Rev. B {\bf 58}, 5719 (1998).
\bibitem{Fong}H. F. Fong {\em et al.}, preprint, cond-mat/9910041; see also, B. Keimer
{\em et al.}, Physica C {\bf 282-287}, 232 (1997).
\bibitem{Hanke}F. Ronning {\em et al.}, Science {\bf 282}, 2067 (1998).
\bibitem{Marston}J. B. Marston and I. Affleck, Phys. Rev. B {\bf 39}, 11 538
(1989).
\bibitem{Laughlin}R. B. Laughlin, Phys. Rev. Lett. {\bf 79}, 1726 (1997).
\bibitem{Batlogg}See, for example, B. Batlogg in {\em High Temperature
Superconductivity}, edited by K. Bedell {\em et al.}, 37 (Addison-Wesley, Redwood City,
1990).
\bibitem{Ising}See, for example, D. J. Thouless, Phys.
Rev. {\bf 187}, 732 (1969).
\bibitem{Laughlin2}R. B. Laughlin, Adv. in Phys. {\bf 47}, 943 (1998).
\bibitem{Chakravarty}S. Chakravarty and R. B. Laughlin, unpublished.
\bibitem{Yin}L. Yin, M. Troyer, and S. Chakravarty, Europhys. Lett. {\bf 42}, 559
(1998).
\bibitem{footnote} The even-odd response functions are interchanged in the
$\sigma$-model because the unit vector field in this model is related to the
spin operator $(-1)^{i+p}{\bf S}_i^{(p)}$, where $i$ are the sites  of the
$2D$ lattice and $p=1,2$ are indices of the layers within a
bilayer\cite{Yin}.
\end{references}
\end{document}